\documentclass{article}
\usepackage{frascatiphys,graphicx,amsfonts,amsmath,amssymb,slashed,xcolor,hyperref}
\hypersetup{colorlinks=true,linkcolor=blue,citecolor=magenta,filecolor=magenta, urlcolor=cyan}
\newcommand{\be}{\begin{equation}}
\newcommand{\ee}{\end{equation}}
\begin{document}
\title{STRONG DYNAMICS \& DARK MATTER: INVESTIGATING A MINIMAL SETUP}
\author{
  Benjamin Fuks\\
    {\em Laboratoire de Physique Th\'eorique et Hautes Energies (LPTHE), UMR 7589,
         Sorbonne Universit\'e et CNRS, }\\
    {\em 4 place Jussieu, 75252 Paris Cedex 05, France}\\
    {\em Institut Universitaire de France, 103 boulevard Saint-Michel,  75005 Paris, France}\\[.2cm]
  Federica Giacchino \\
    {\em INFN, Laboratori Nazionali di Frascati, C.P. 13, 100044, Frascati, Italy} \\[.2cm]
  Laura Lopez-Honorez \\
    {\em Service de Physique Th\'eorique, CP225, Universit\'e Libre de Bruxelles, Bld du Triomphe,
         1050 Brussels, Belgium}\\
    {\em  Vrije Universiteit Brussel and The International Solvay Institutes, Pleinlaan 2,
          1050 Brussels, Belgium}\\[.2cm]
  Michel H.G. Tytgat and  J\'er\^ome Vandecasteele\\
    {\em Service de Physique Th\'eorique, CP225, Universit\'e Libre de Bruxelles, Bld du Triomphe,
    1050 Brussels, Belgium}
}
\maketitle
\baselineskip=10pt
\begin{abstract}
  We discuss the phenomenology of a dark matter scenario in which
  we extend the Standard Model by a real scalar particle and a vector-like heavy
  quark. Such a model can be seen as a simplified version of a
  composite setup in which the scalar field, that couples to the
  top quark via a Yukawa interaction with the new heavy quark, is a viable dark
  matter candidate. We emphasize that QCD corrections are important not only for
  predictions at colliders but also for direct and indirect dark matter searches
  and the relic abundance. We moreover show that a large
  fraction of the model parameter space remains unconstrained.
\end{abstract}

\baselineskip=14pt
\section{Introduction}

There is  a large experimental effort worldwide that aims at deciphering the nature of the dark matter. A much studied scenario assumes that the dark matter (DM) is made of a stable and neutral particle species, with a relic abundance fixed by chemical freeze-out in the early universe. As the required annihilation cross section is in the 1~pb range, such particles are collectively called  weakly interacting massive particles (WIMPs). A generic feature of any DM candidate with an abundance originating from freeze-out is that they can be complementarily searched for directly, indirectly and at colliders.
We report on a phenomenological analysis of a simple, yet rich, WIMP model in which DM is a real scalar particle that couples dominantly to the top quark\cite{Colucci:2018vxz}. Such a simplified scenario could seen as the dark sector of more ambitious
theories beyond the Standard Model (SM), like non-minimal composite models\cite{Cacciapaglia:2019ixa}.

\section{Theoretical context}

The Lagrangian describing our simplified model takes the form
\be
 {\cal L} = {\cal L}_{\rm SM} + i \bar T \slashed{D} T - m_T \bar T T
      +  \frac 12 \partial_\mu S \partial^\mu S - \frac 12 m_S S^2
      + \Big[ \tilde{y}_t\, S\ \bar T\,  t_R + {\rm h.c.} \Big] \ ,
\label{eq:model}\ee
where $S$ denotes the scalar dark matter candidate, $T$ is a vector-like color triplet fermion and $t_R$ stands for the right-handed top quark. Other terms are forbidden by imposing that both $S$ and $T$ are odd under a $Z_2$ symmetry whereas the SM fields are set to be even. This ensures the stability of the dark matter and forbids the mixing of the $T$-quark with the SM quarks.
Whilst in practice the quantum numbers of the $T$ particle also allow for
couplings with $S$ and the first and second generation quarks,
 we set these to zero (the associated phenomenology having been worked out in the past\cite{Giacchino:2015hvk}), thus assuming that DM dominantly interacts with the
top quark.
Non-minimal composite models, in which the top quark plays a special role\cite{Cacciapaglia:2019ixa}, could yield the Lagrangian of eq.~\eqref{eq:model}. While such a possibility is very much worth further investigating, we consider in the meantime this Lagrangian as a simplified model\cite{Abdallah:2015ter}. We so assume that only 3 parameters are needed to study its phenomenology, namely the two new physics masses $m_S$ and $m_T$, and the Yukawa coupling $\tilde{y}_t$.

\begin{figure}
  \centering
  \includegraphics[height=6cm]{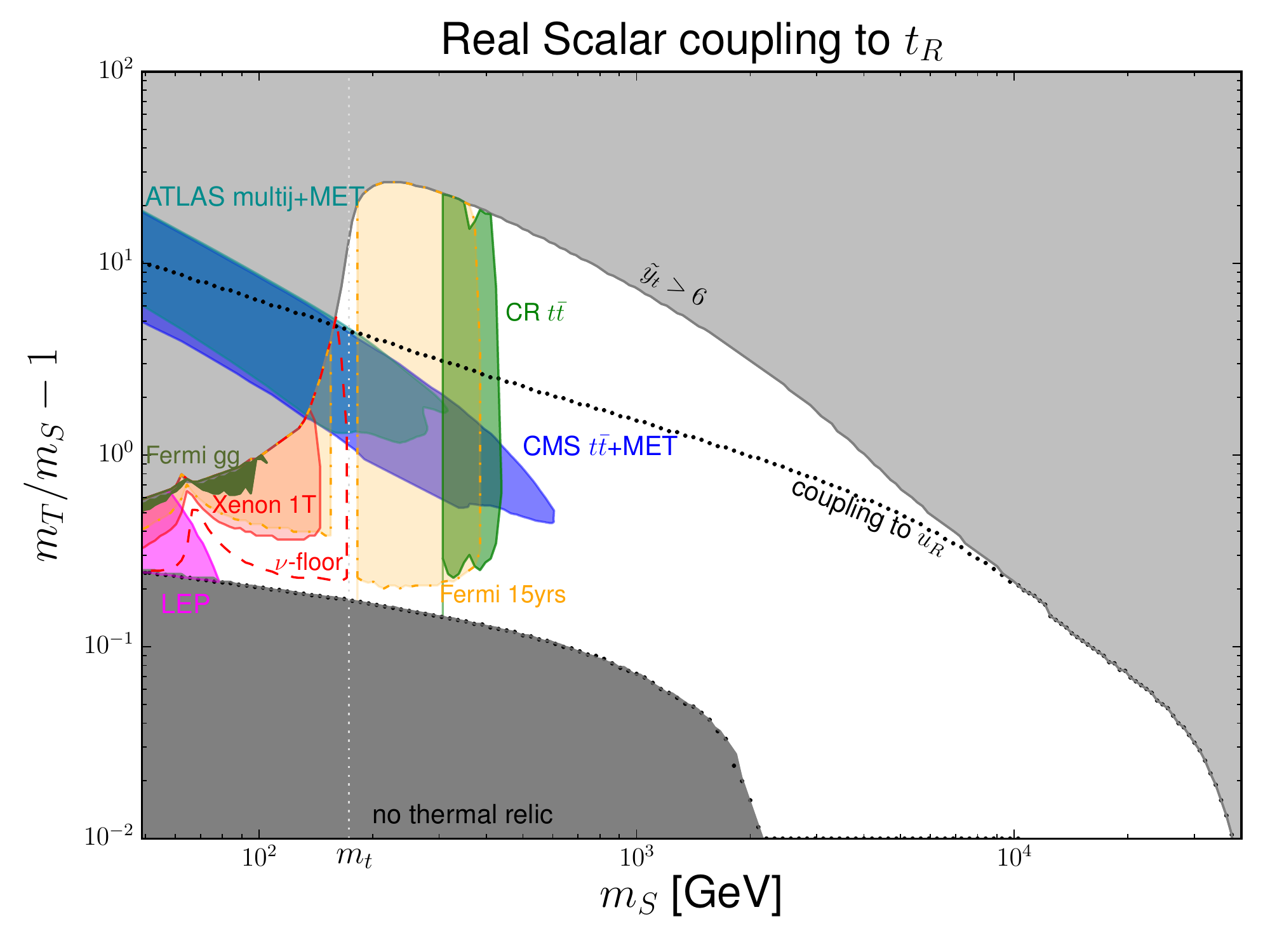}
  \caption{Top-philic DM model parameter space shown in the DM mass ($m_S$) and
    spectrum compression factor ($m_T/m_S-1$) plane. In the gray regions, the
    observed DM relic density cannot be accommodated, whilst in the
    region in between, there exists a specific $\tilde y_t$ value leading to the right DM
    abundance. We refer to the text for the description of the different
    experimental constraints, that each corresponds to a given colored region.
    We moreover impose $\tilde y_t < 6$ to allow for a perturbative treatment
    in our calculations (upper gray region). In the lower gray region, DM is
    under-abundant assuming thermal freeze-out.}
\label{Fig::CCL-viab}
\end{figure}

In the sequel, we first summarize our determination of the relic abundance and then discuss the resulting experimental constraints on the model parameters. We put a special emphasis on the role of the QCD radiative corrections, which are particularly important in our model. The results of our analysis including DM direct and indirect detection, as well as the bounds stemming from LHC searches, are collected in figure~\ref{Fig::CCL-viab}. This exhibits the complementarity between the experimental searches, and that heavy DM configurations are untested.

\section{Relic density}
\label{sec:relic}

\begin{figure}
  \centering
  \includegraphics[height=5cm]{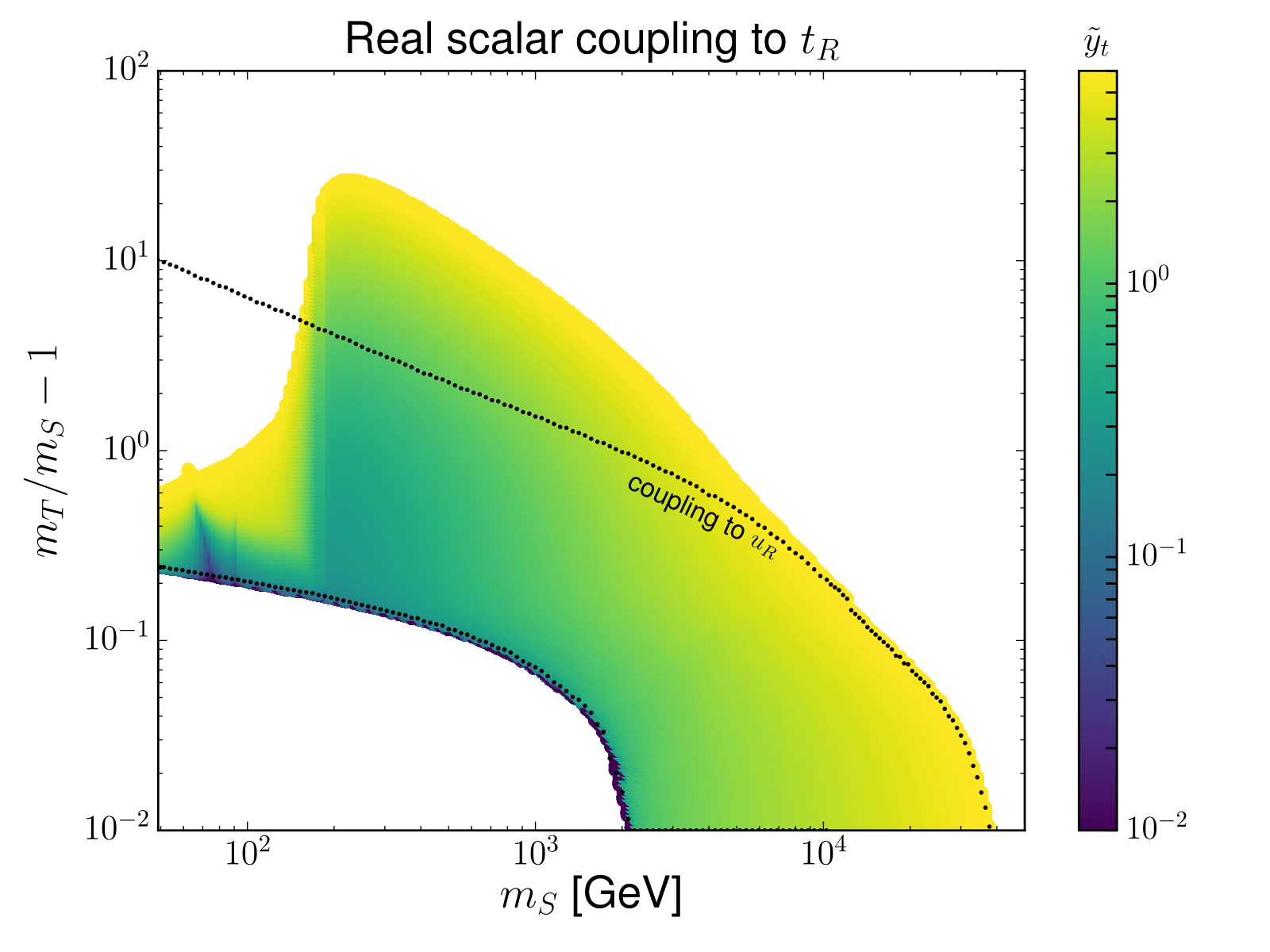}
  \includegraphics[height=5cm]{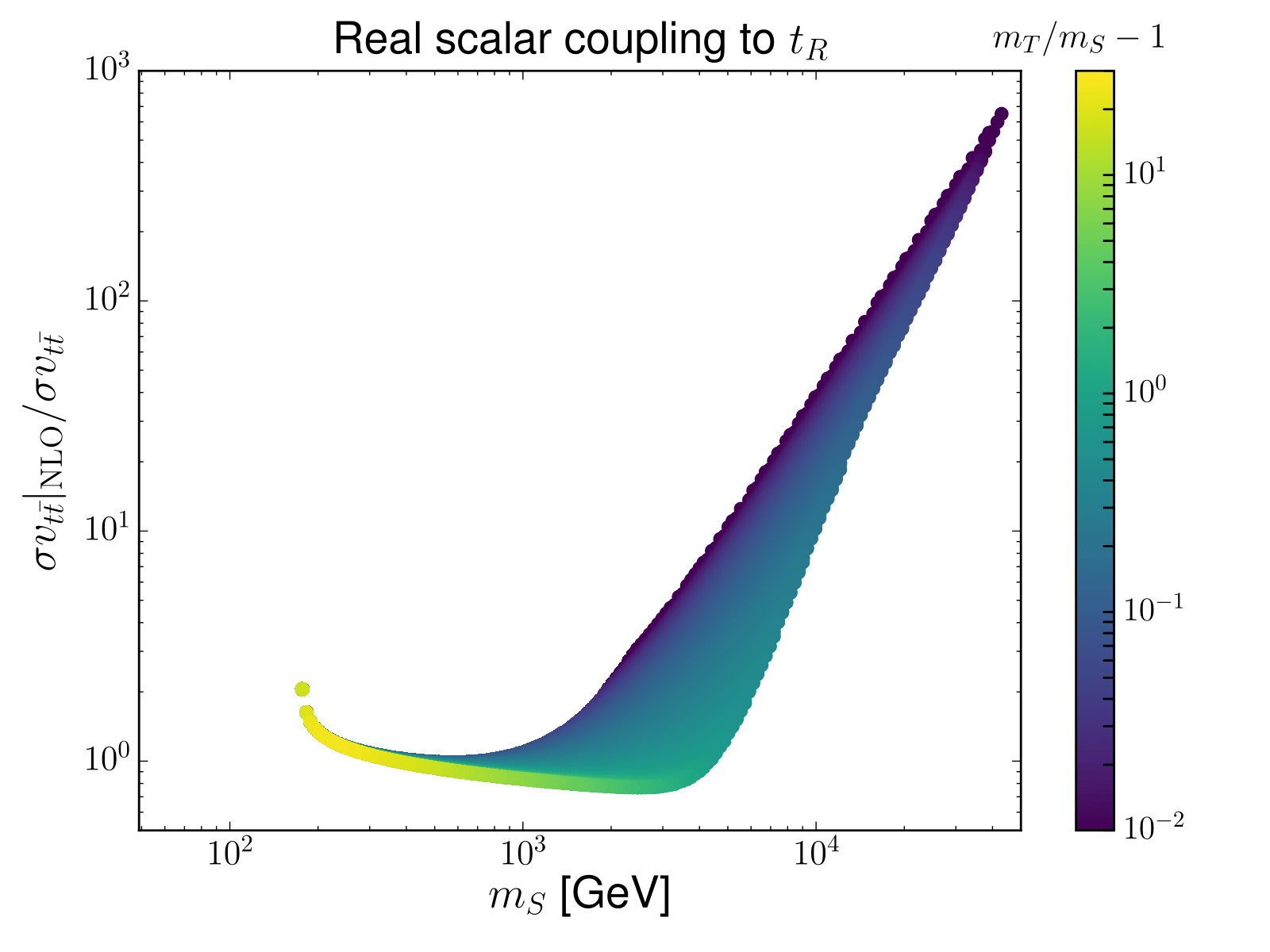}
  \caption{{Left}: Parameter space region in which the observed DM abundance,
   $\Omega h^2=0.12$, can be accommodated. The results are presented in the
   $(m_S, m_T/m_S-1)$ plane and the color gradient refers to the corresponding
   $\tilde{y}_{t}$ value. The dashed line refers to the bound obtained when
   neglecting the top-quark mass. {Right}: Ratio of the
   thermally-averaged NLO annihilation cross section to
   the LO one for each viable scenario.
   The color gradient represents the compression parameter.}
\label{Fig::relic-m-r}
\end{figure}

Assuming thermal freeze-out, all viable setups for which the DM relic
abundance can match the Planck collaboration results correspond to the
area in between the gray regions in figure~\ref{Fig::CCL-viab}, or
equivalently, to the colored region in the left panel of
figure~\ref{Fig::relic-m-r}. Both results are depicted in the plane
$(m_S, m_T/m_S-1)$ where we coin $m_T/m_S-1$ the spectrum compression
factor as it shows how close are the mediator and DM masses. It turns
out that a viable DM candidate can be continuously obtained from
masses ranging from a few GeV to up to 40~TeV. While not unexpected
for WIMP candidates, this parameter range is quite large, owing to the
various possible annihilation channels.  For $m_S\gtrsim 5$~TeV ({\it
  i.e.}, $m_S\gg m_t$), the dominant annihilation channel involves
additional QCD radiation, $SS\to t\overline{t}g$. This originates from
the $d$-wave suppression of annihilations into fermion pairs in the
$m_S \gg m_f$ limit and from a strong enhancement of the so-called
virtual internal bremsstrahlung
contributions\cite{Toma:2013bka,Giacchino:2013bta}. The latter is
illustrated in the right panel of figure~\ref{Fig::relic-m-r} in which
we present the ratio of the next-to-leading-order (NLO) annihilation
cross section into a $t\bar t$ pair (including extra gluon emission)
to the leading-order (LO) one at freeze-out time. For $m_S\gg m_t$,
this enhancement is significant and the NLO contributions clearly
dominate. As $m_S$ decreases the top quark mass becomes less
negligible, and, while NLO effects remain important, the ratio between
the NLO and LO predictions gets closer to 1. Finally, the apparent
increase at $m_S\sim m_t$ is spurious and should be removed by a
proper treatment of the threshold
effects\cite{Colucci:2018qml,Bringmann:2015cpa}.  On the left panel of
figure~\ref{Fig::relic-m-r}, the impact of a non-negligible top quark
mass can be seen by comparing the colored region associated with an $S$ coupling
to $t_R$ to the viable parameter space region when $S$ couples to the right-handed
up quark$u_R$\cite{Giacchino:2015hvk} (shown in between the dotted black lines). For
$m_S\leq m_t$, the relic density could arise from loop-induced $SS\to
gg$ annihilations\cite{Giacchino:2015hvk,Giacchino:2014moa}, a process
that is also unexpectedly large if the mediator is not too heavy ($m_T
\gtrsim m_S$) and the compression factor close to 1.

The abundance may also originate from co-annihilations ({\it e.g.}, $S
T \to g t$) or even from mediator annihilation $T\bar T \to g g/q\bar q$ if the mass
spectrum is sufficiently compressed (typically in the dark blue region
of the left panel of figure~\ref{Fig::relic-m-r} for $m_S\lesssim 3$ TeV), provided the $S$ and
$T$ particles are in chemical equilibrium ($\Gamma(S \leftrightarrow
T) \gtrsim H$ with $H$ being the Hubble rate). The rationale for such a compressed
mass spectrum in which a DM particle is degenerate with colored states
in a natural way may stem from extra-dimensional\cite{Hooper:2007qk}
or grand unified\cite{Ferrari:2018rey} theories. In addition,
departures from thermal equilibrium are known to potentially affect the
results\cite{Garny:2018icg}, and while we could
expect that Sommerfeld corrections strongly impact the $T \bar T \to g
g/q \bar q$ annihilation cross sections, the existence of both
attractive and repulsive channels tame those effects that are at most
of ${\cal O}(15 \%)$\cite{Colucci:2018vxz,Giacchino:2015hvk}.

\section{Experimental and observational constraints}
\subsection{LHC searches}

\begin{figure}
\centering
  \includegraphics[height=5cm]{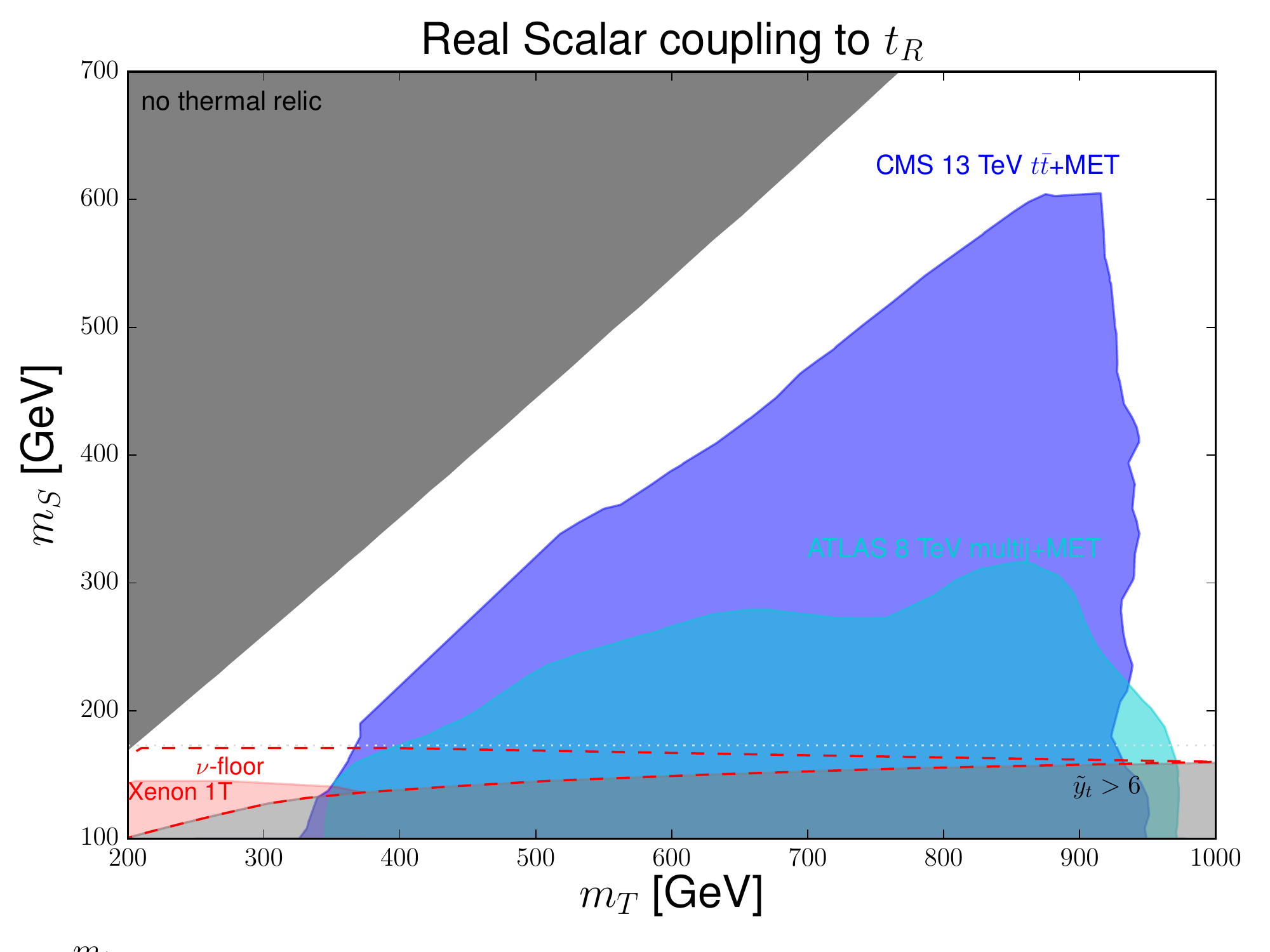}
  \includegraphics[height=5cm]{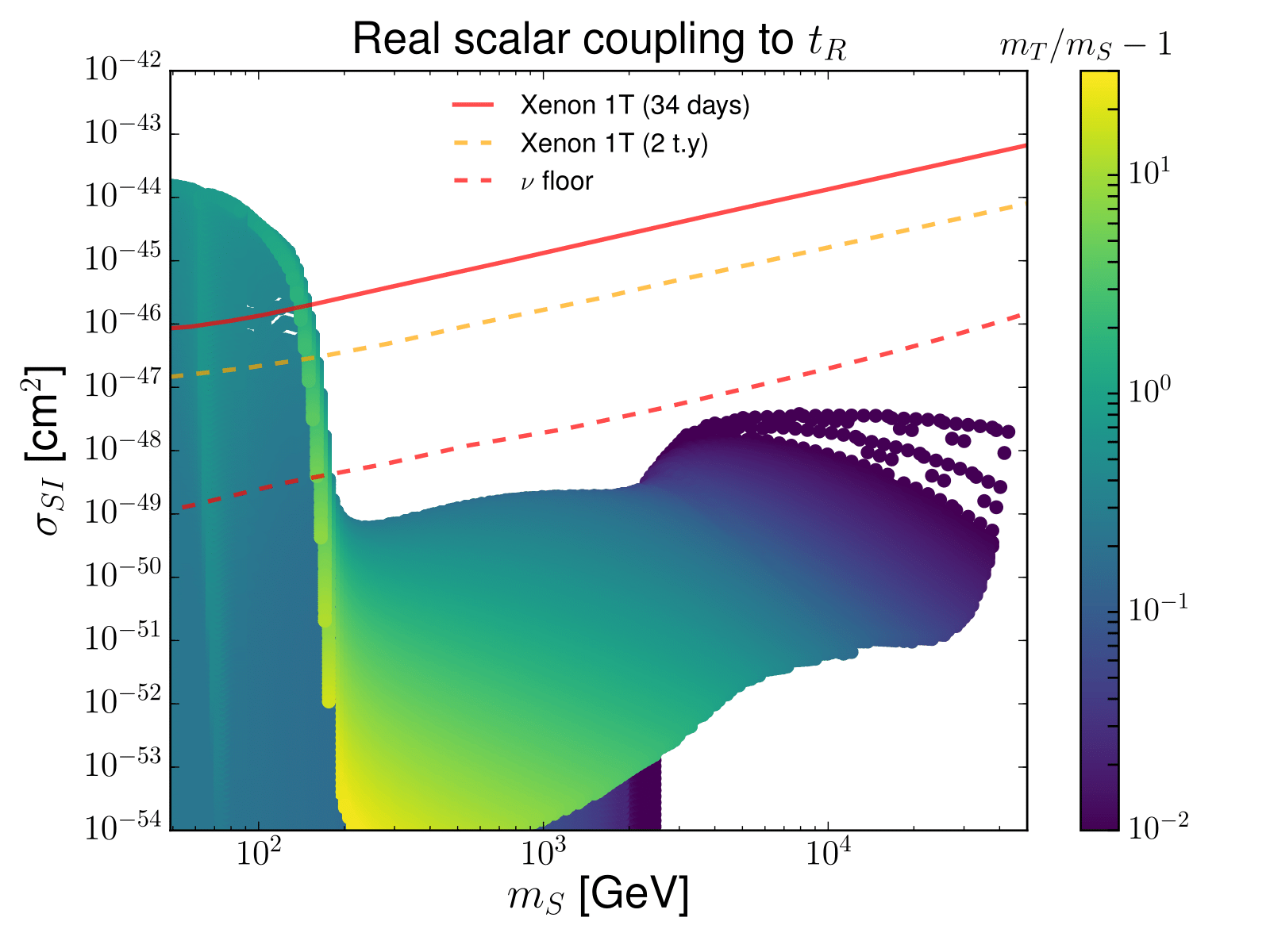}
\caption{Left: LHC constraints on the model expressed in the $(m_S, m_T)$   mass plane, together with the DM relic density and direct detection bounds. Right: Spin-independent DM-nucleon scattering cross section as a function of $m_S$, for each scenario accommodating Planck data. The compression factor is depicted by the color code, and we superimpose current (solid red) and future (dashed orange) 90\% confidence level exclusions from Xenon 1T.
}
\label{Fig::DD-m-SI}
\end{figure}

Like for any WIMP-like DM, our model can be probed at the LHC through
signatures comprised of missing transverse energy (MET) produced in
association with either jets (mono-X-like probes) or a $t\bar t$
pair. We reinterpret the results of a typical DM search in the $t\bar
t$ plus MET mode using 35.9~fb$^{-1}$ of CMS
data\cite{Sirunyan:2017leh}, increasing sensitivity to compressed
scenarios by additionally considering a dedicated CMS
search\cite{CMS:2017odo}. We moreover reinterpret the results of two
early Run~2 ATLAS DM searches in the monojet and multijet plus MET
modes\cite{Aaboud:2016tnv,Aaboud:2016zdn}. Those searches being
limited by systematics, any constraint they could lead to is not
expected to get more severe with more data\cite{Banerjee:2017wxi}.
Our results are presented in the left panel of
figure~\ref{Fig::DD-m-SI}. The colored regions correspond to scenarios
excluded at the 95\% confidence level by at least one of the
considered $t\bar t$ plus MET (dark blue) or multijet plus MET (light
blue) analyses, considering NLO simulations for the DM signal. The
latter dominantly stems from the production of a pair of $T$-quarks
decaying into top quarks and missing energy ($pp\to T\bar T\to tS\bar
t S$) and is excluded for $m_T$ values lying in the 300--1000~GeV
range, provided there is enough phase space to guarantee the mediator
decay. No constraint arises if the $T\to t S$ decay channel is closed,
as the $T$-quark turns out to be long-lived. Those results are
reported according to the same color code in
figure~\ref{Fig::CCL-viab}.

\subsection{Direct detection}

DM being scalar, direct detection constraints can only originate from spin-independent exclusion limits imposed by ton-size liquid Xenon experiments (currently Xenon~1T\cite{Aprile:2015uzo,Aprile:2017iyp}).
DM-nucleon scattering occurring at one loop through the exchange of virtual top quarks, the coupling between $S$ and nucleons boils down to an $SSgg$ effective operator. This contrasts with models in which DM couples to light quarks, where higher-twist operators and long-range interactions are important\cite{Giacchino:2015hvk}.
The constraints on the scattering cross-section are presented in the right panel of figure~\ref{Fig::DD-m-SI}, the strongest bounds arising for light DM candidates. In this case, the relic density is typically driven by annihilations into gluons ($SS\rightarrow gg$) mediated by a large $\tilde y_t$ Yukawa coupling. This suggests a potentially large value for the DM-nucleon scattering cross section. For heavier dark matter, however the scattering cross section typically lies below the neutrino floor (red dashed). This severely limits the relevance of DM direct detection searches for setups like the one of eq.~\eqref{eq:model}. The state of affair is reported as the red colored area in figure~\ref{Fig::CCL-viab}.

\subsection{Indirect detection}
\begin{figure}
  \centering
  \includegraphics[height=5cm]{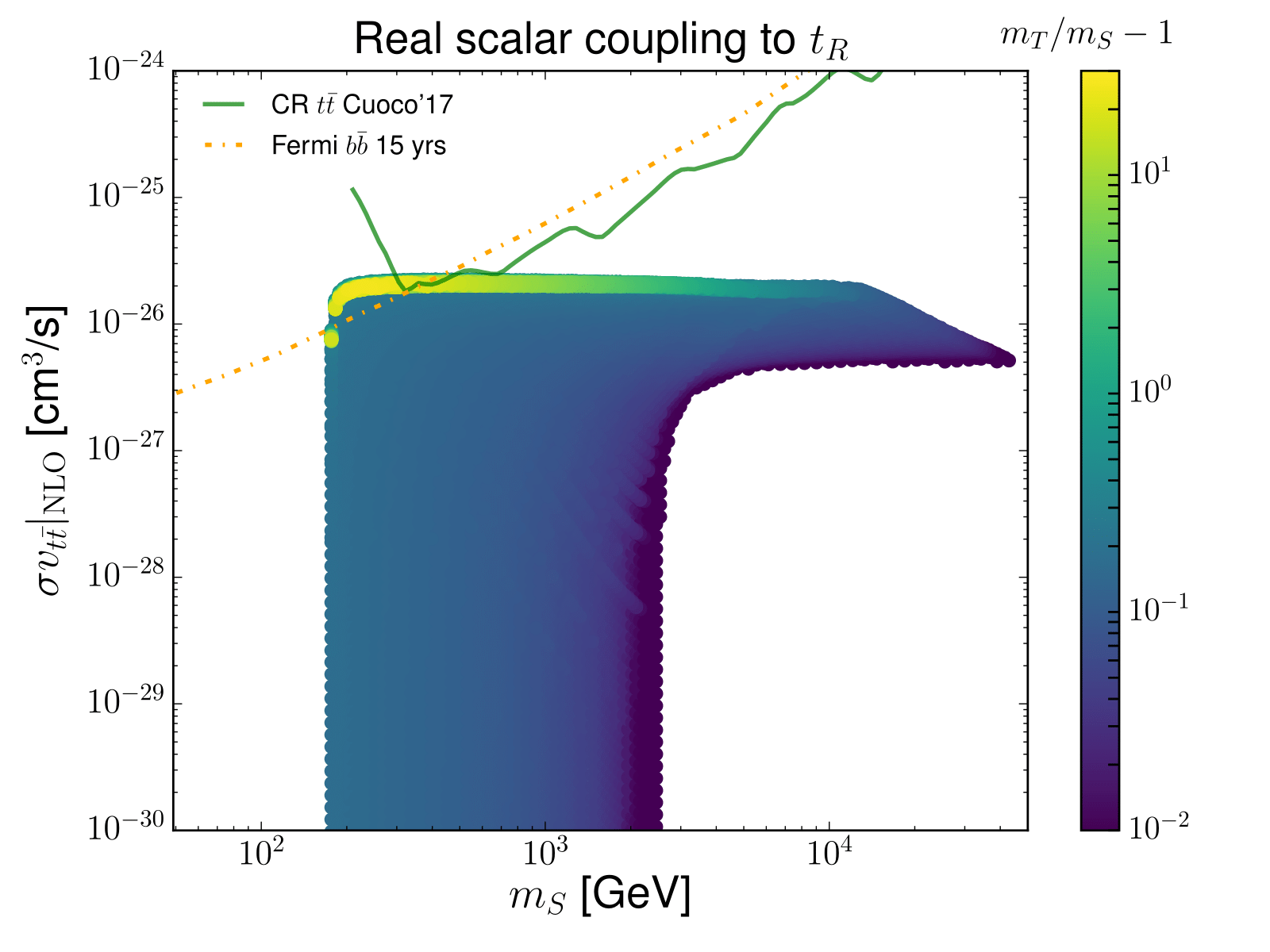}
  \includegraphics[height=5cm]{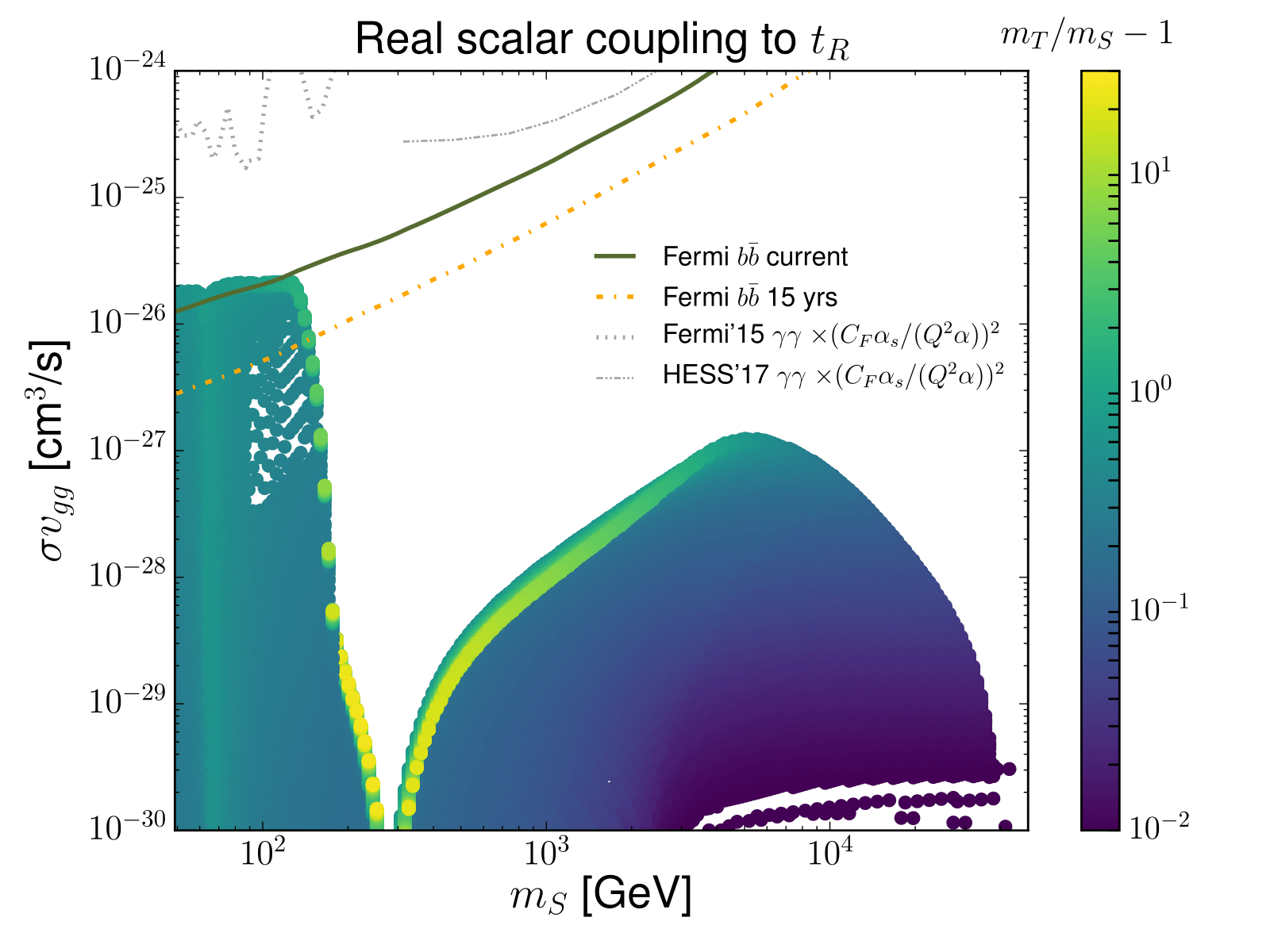}
  \caption{NLO $SS \to t\bar t$ (left) and loop-induced LO $SS\to gg$ (right)
    annihilation cross sections at zero velocity, relevant for indirect
    searches. We superimpose limits from antiproton cosmic rays (solid light
    green) and from current (solid dark green) and future (dot-dashed orange)
    Fermi-LAT dwarf spheroidal galaxy data in the $b\bar b$ channel (after an
    appropriate recasting).
}
\label{Fig::ID-m-sv}
\end{figure}

DM annihilations into $gg$ or $t\bar{t} (g)$ systems would produce a continuum of gamma rays and cosmic rays (in particular antiprotons). Of particular relevance for indirect detection is the effect of bremsstrahlung of gluons, related to the issue of disentangling hard and soft gluon emission to control the associated infrared divergences\cite{Colucci:2018qml,Bringmann:2015cpa}.
As shown in figure~\ref{Fig::ID-m-sv}, some model configurations are excluded by current indirect detection searches\cite{Cuoco:2017iax,Charles:2016pgz,Rinchiuso:2017kfn}. In the right panel of the figure, we present the constraints arising from annihilations into gluons pairs, while in the left panel, we consider annihilations into a $t \bar t (g)$ final state. The former is most relevant for lighter DM, $m_S \lesssim 100$~GeV, some scenarios being excluded.
Annihilations into the $t\bar t (g)$ mode can yield constraints from Fermi-LAT dwarf galaxy results. While assuming a $b\bar b$ final state, the latter can be recasted\cite{Colucci:2018vxz}. Some model configurations for which $m_t<m_S<500$ GeV turn out to be excluded.
Finally, DM annihilations can feature gamma-ray-line topologies to which experiments like Fermi-LAT are very sensitive to. These turn out to be subdominant compared with the gamma-ray continuum generated by the hadronization of the $t \bar t$ decay products and gluons (right panel). All indirect detection constraints are reported in figure~\ref{Fig::CCL-viab} following the same color coding as in figure~\ref{Fig::ID-m-sv}.

\section{Summary}

WIMP dark matter is being tested in various experiments in astrophysics,
cosmology and at colliders. In this work, we have extensively investigated a
simplified top-philic scalar DM scenario that could find its origin in composite
setups. We have studied various existing constraints on the model and shown
that although there is a complementarity between the different searches, only a
small fraction of the viable parameter space is currently tested. The most
fruitful long-term strategy therefore consists in an increase of the energy reach
at colliders.

\section{Acknowledgements}
We thank Stefano Colucci for collaboration on the works reported in these proceedings. This  study  has  been  partly  supported  by French  state  funds  managed  by  the  Agence  Nationale de  la  Recherche  (ANR)  in  the  context  of  the  LABEXILP (ANR-11-IDEX-0004-02, ANR-10-LABX-63), by the FRIA, the FNRS, the Strategic Research Program \textit{High Energy Physics} and  the  Research  Council  of  the  Vrije Universiteit Brussel, the MIS research grant number F.4520.19, the IISN convention 4.4503.15 and by the Excellence of Science (EoS) convention 30820817.


\begin{thebibliography}{99}

\bibitem{Colucci:2018vxz}
  S.~Colucci, B.~Fuks, F.~Giacchino, L.~Lopez Honorez, M.~H.~G.~Tytgat and J.~Vandecasteele,
  Phys.\ Rev.\ D {\bf 98} (2018) 035002.
  

\bibitem{Cacciapaglia:2019ixa}
  G.~Cacciapaglia, H.~Cai, A.~Deandrea and A.~Kushwaha,
  JHEP {\bf 1910} (2019) 035.

\bibitem{Giacchino:2015hvk}
  F.~Giacchino, A.~Ibarra, L.~Lopez Honorez, M.~H.~G.~Tytgat and S.~Wild,
  JCAP {\bf 1602} (2016) 002.


\bibitem{Abdallah:2015ter}
  J.~Abdallah {\it et al.},
  Phys.\ Dark Univ.\  {\bf 9-10} (2015) 8.

\bibitem{Toma:2013bka}
  T.~Toma,
  Phys.\ Rev.\ Lett.\  {\bf 111} (2013) 091301.

\bibitem{Giacchino:2013bta}
  F.~Giacchino, L.~Lopez-Honorez and M.~H.~G.~Tytgat,
  JCAP {\bf 1310} (2013) 025.
  
\bibitem{Colucci:2018qml}
  S.~Colucci, F.~Giacchino, M.~H.~G.~Tytgat and J.~Vandecasteele,
  Phys.\ Rev.\ D {\bf 98} (2018) 115029.

\bibitem{Bringmann:2015cpa}
  T.~Bringmann, A.~J.~Galea and P.~Walia,
  Phys.\ Rev.\ D {\bf 93} (2016) 043529.

\bibitem{Giacchino:2014moa}
  F.~Giacchino, L.~Lopez-Honorez and M.~H.~G.~Tytgat,
  JCAP {\bf 1408} (2014) 046.



\bibitem{Hooper:2007qk}
  D.~Hooper and S.~Profumo,
  Phys.\ Rept.\  {\bf 453} (2007) 29.

\bibitem{Ferrari:2018rey}
  S.~Ferrari, T.~Hambye, J.~Heeck and M.~H.~G.~Tytgat,
  Phys.\ Rev.\ D {\bf 99} (2019) 055032.

\bibitem{Garny:2018icg} 
  M.~Garny, J.~Heisig, M.~Hufnagel and B.~Lülf,
  Phys.\ Rev.\ D {\bf 97} (2018) 075002.


\bibitem{Sirunyan:2017leh}
  A.~M.~Sirunyan {\it et al.} [CMS Collaboration],
  Phys.\ Rev.\ D {\bf 97} (2018) no.3,  032009.

\bibitem{CMS:2017odo}
  CMS Collaboration,
  CMS-PAS-SUS-16-052.

\bibitem{Aaboud:2016tnv}
  M.~Aaboud {\it et al.} [ATLAS Collaboration],
  Phys.\ Rev.\ D {\bf 94} (2016) 032005.

\bibitem{Aaboud:2016zdn}
  M.~Aaboud {\it et al.} [ATLAS Collaboration],
  Eur.\ Phys.\ J.\ C {\bf 76} (2016) 392.

\bibitem{Banerjee:2017wxi}
  S.~Banerjee, D.~Barducci, G.~Bélanger, B.~Fuks, A.~Goudelis and B.~Zaldivar,
  JHEP {\bf 1707} (2017) 080.



\bibitem{Aprile:2015uzo}
  E.~Aprile {\it et al.} [XENON Collaboration],
  JCAP {\bf 1604} (2016) 027.

\bibitem{Aprile:2017iyp}
  E.~Aprile {\it et al.} [XENON Collaboration],
  Phys.\ Rev.\ Lett.\  {\bf 119} (2017) 181301.

\bibitem{Cuoco:2017iax}
  A.~Cuoco, J.~Heisig, M.~Korsmeier and M.~Krämer,
  JCAP {\bf 1804} (2018) 004.

\bibitem{Charles:2016pgz}
  E.~Charles {\it et al.} [Fermi-LAT Collaboration],
  Phys.\ Rept.\  {\bf 636} (2016) 1.

\bibitem{Rinchiuso:2017kfn}
  L.~Rinchiuso {\it et al.} [HESS Collaboration],
  PoS ICRC {\bf 2017} (2018) 893.

\end{thebibliography}
\end{document}